\documentclass[pdflatex,sn-mathphys-ay]{sn-jnl}

\usepackage{graphicx}%
\usepackage{subfig}
\usepackage{multirow}%
\usepackage{amsmath,amssymb,amsfonts}%
\usepackage{amsthm}%
\usepackage{mathrsfs}%
\usepackage[title]{appendix}%
\usepackage{xcolor}%
\usepackage{textcomp}%
\usepackage{manyfoot}%
\usepackage{booktabs}%
\usepackage{algorithm}%
\usepackage{algorithmicx}%
\usepackage{algpseudocode}%
\usepackage{listings}%

\theoremstyle{thmstyleone}%
\theoremstyle{thmstyletwo}%

\theoremstyle{thmstylethree}%

\begin{document}

\title[Article Title]{Model Predictive Controller to Regulate Cortisol Levels in Individuals With Adrenal Insufficiency }


\author*[1]{\fnm{Renuka} \sur{Joshi}}\email{joshi278@purdue.edu}

\author[1]{\fnm{Nayana} \sur{Saha}}\email{saha80@purdue.edu}


\author[1]{\fnm{Vittal} \sur{Srinivasan}}\email{srini133@purdue.edu}

\author[1]{\fnm{Stanislaw H.} \sur{\.{Z}ak}}\email{zak@purdue.edu}

\author[2]{\fnm{Cary N.} \sur{Mariash}}\email{cmariash@iu.edu}

\affil*[1]{\orgdiv{Elmore Family School of Electrical and Computer Engineering}, \orgname{Purdue University}, \orgaddress{\street{610 Purdue Mall}, \city{West Lafayette}, \postcode{47906}, \state{Indiana}, \country{United States}}}

\affil[2]{\orgdiv{Division of Endocrinology}, \orgname{Indiana University School of Medicine}, \orgaddress{\street{340
West 10th Street}, \city{Indianapolis}, \postcode{46202}, \state{Indiana}, \country{United States}}}



\abstract{A model predictive controller (MPC) is used to construct a virtual assistant to aid a physician in prescribing cortisol replacement therapy for patients with adrenal insufficiency (AI). AI, also known as hypocortisolism, is a condition that occurs due to a low concentration of cortisol. This hormonal imbalance significantly impacts the individual's ability to regulate stress, metabolism, and immune responses. Thus, it is essential to maintain cortisol levels within a healthy range. The production of cortisol is governed by the hypothalamus-pituitary-adrenal (HPA) axis, a part of the endocrine system. In this paper, a novel mathematical model of the HPA axis is proposed that incorporates the endogenous circadian rhythm. This model simulates two conditions of hypocortisolism: primary and secondary AI. Adrenal insufficiency cannot be cured, but it can be treated with cortisol replacement therapy. The standard practice is to prescribe a therapeutic dose of hydrocortisone (HC). To evaluate the accuracy of the proposed HPA axis model, an open-loop cortisol replacement strategy with a fixed dosage is used to simulate both primary and secondary AI. The simulation results show that, analytically, it is possible to arrive at a fixed working cortisol replacement strategy. However, this strategy, though effective, is not optimal. To obtain optimal cortisol replacement strategies, an MPC is proposed. An important feature of MPC is that constraints on allowable cortisol replacement dosages can be rigorously addressed. This controller can serve as a virtual assistant to physicians in prescribing daily cortisol replacement therapy.}


\maketitle

\section{Introduction}\label{sec1}

Hypocortisolism, also known as adrenal insufficiency (AI), occurs when the adrenal glands do not produce a sufficient amount of the cortisol hormone. Cortisol is crucial in regulating stress, metabolism, and immune responses. A significant challenge in diagnosing AI is that its symptoms develop gradually over time. Such symptoms include fatigue, low blood pressure, hypoglycemia, muscle weakness, and reduced body immunity to infections. In addition, there are psychological effects such as depression, anxiety, and reduced resilience to stress. Chronic stress is a primary factor contributing to the dysregulation of the hypothalamic-pituitary-adrenal (HPA) axis. This hormonal imbalance is related to conditions such as Addison's disease and hypothalamic-pituitary (HP) disorder, which are modeled in this paper. The cycle of chronic stress and cortisol deficiency continues to weaken the ability to cope with external stress. Research on the effects of stress-related hypocortisolemic disorders can lead to early detection and targeted therapies. Therefore, it is essential to address this condition to reduce its physical, psychological, and social impacts by advancing the understanding of stress-related health disorders.

\begin{figure*}
    \centering
    \includegraphics[width=0.5\textwidth]{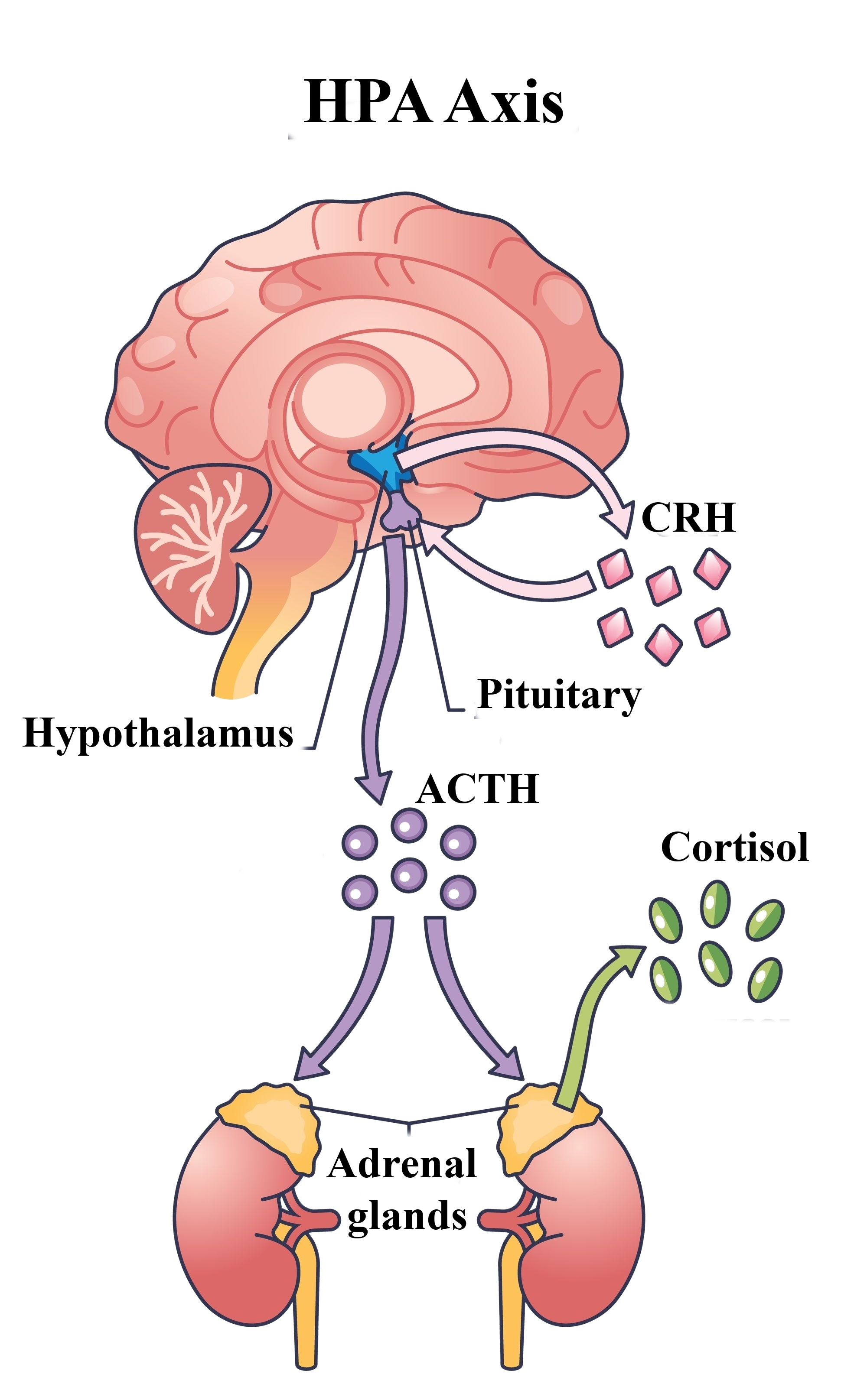}
    \caption{Hypothalamus-Pituitary-Adrenal (HPA) axis (Source: Shutterstock).}
    \label{fig:HPA-axis}
\end{figure*}

The HPA axis, shown in Fig.~\ref{fig:HPA-axis}, regulates stress and immune system functions. It consists of hypothalamus, a part of the limbic system located beneath the thalamus in the midbrain; the pituitary gland, situated just below the hypothalamus at the base of the brain; and the adrenal gland, located on top of the kidneys. This axis is responsible for responding to stress. It secretes three significant hormones: (i) Corticotropin-Releasing Hormone (CRH), (ii) Adrenocorticotropic Hormone (ACTH), and (iii) Cortisol. In response to stress, the hypothalamus releases CRH from the paraventricular nucleus (PVN), which, in turn, stimulates the anterior pituitary gland to secrete ACTH. ACTH then travels through the bloodstream to the adrenal glands, where it causes the release of cortisol. The HPA axis has feedback loops that inhibit the production of elevated CRH and ACTH. 

Circadian rhythm, along with feedback loops, plays an important role in the functioning of the HPA axis, as noted by~\cite{https://doi.org/10.1002/wsbm.1518}. Both the circadian rhythm and feedback loops synchronize the release of cortisol with diurnal variations. The oscillatory behavior of the circadian function is due to negative feedback loops from cortisol to CRH and ACTH. The role of the circadian rhythm and the feedback loops is to maintain homeostasis when the body is exposed to external or internal stress.

\subsection{Literature Review}
\label{Sec:literature review}
In a healthy individual, cortisol levels vary throughout the day in response to the circadian rhythm. The cortisol-awakening response (CAR) is characterized by elevated cortisol levels in the morning, as noted in \cite{Kudielka01012010}. As suggested by~\cite{HOSSEINICHIMEH201552} and~\cite{https://doi.org/10.1111/j.1600-0447.2007.00967.x}, cortisol levels are typically low from 8:00~pm to 4:00~am, peak between 4:00~am and 12:00~pm, and then decline back to minimal levels from 12:00~pm until 8:00~pm.

The normal range of CRH is 2--28 pg/mL, ACTH is 5--50 pg/mL, and cortisol is 3--25 $\mu$g/dL,---see, for example,~\cite{10.1210/jcem-64-5-1047} and~\cite{cortisolACTH2023}. The hormone ranges could vary as per age, gender, stress level, history of trauma, weight, smoking status, alcohol or substance abuse, sleep duration, and menstrual cycle, as discussed in~\cite{SPEER2019100180}. High cortisol levels lead to hypercortisolism, while low cortisol levels cause AI.

\begin{figure}
\centerline{\includegraphics[width=0.5\textwidth]{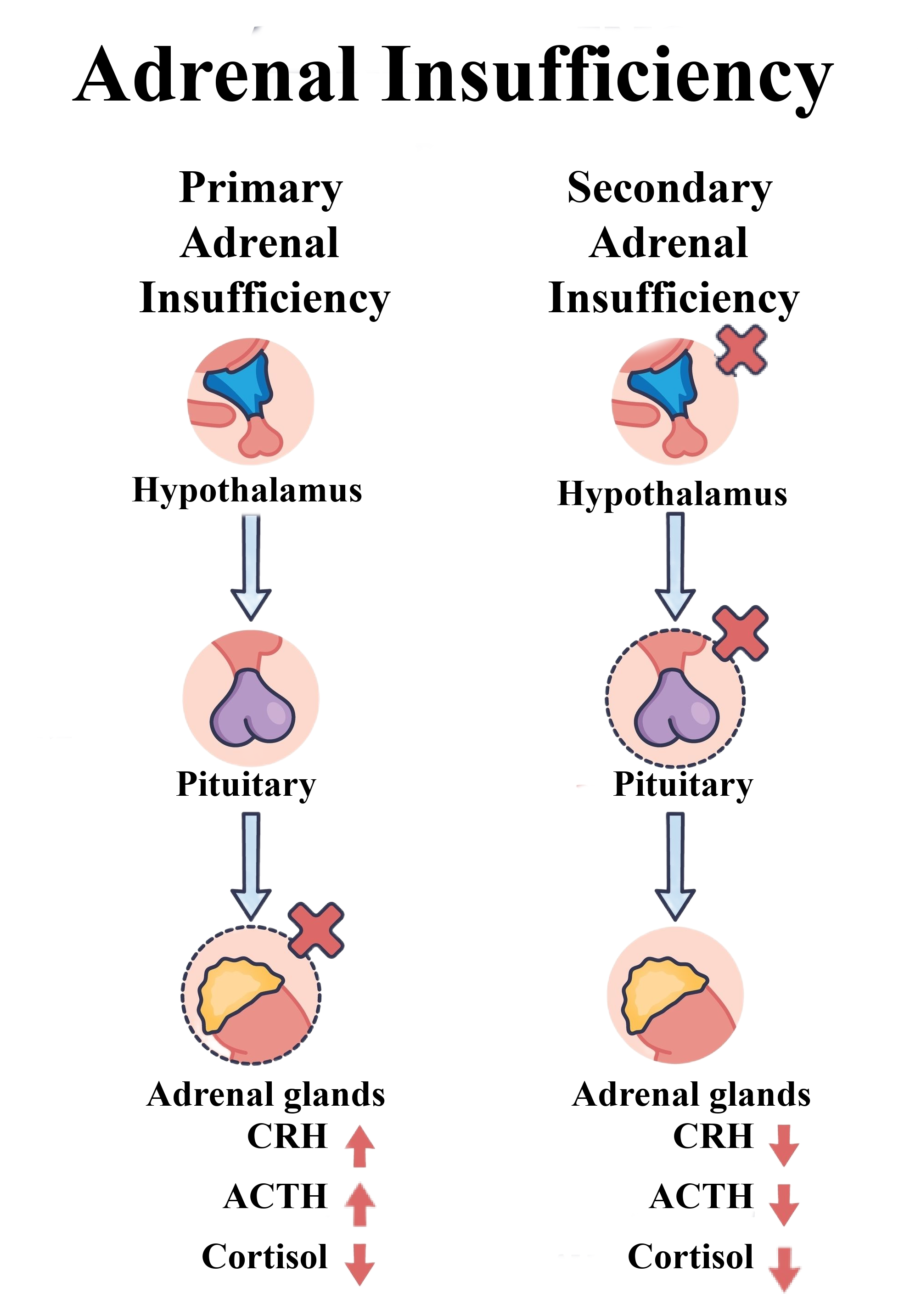}}
\caption{Types of adrenal insufficiency (Source: Modified from Shutterstock).}
\label{fig:AI}
\end{figure}

Our paper focuses on two conditions of AI, as illustrated in Fig.~\ref{fig:AI}. In secondary AI, the glands are unable to produce adequate amounts of CRH and ACTH hormones. As a result, the adrenal glands cannot produce sufficient cortisol. In primary AI, levels of CRH and ACTH are elevated due to the adrenal glands' inability to produce an adequate amount of cortisol. AI can be diagnosed by the ACTH stimulation test (also known as the Cosyntropin test), in~\cite{Hypocortisolism}. Although there is no permanent cure for AI, the condition is generally regulated with appropriate medication. Standard prescription for AI patients is cortisol replacement therapy taken three times a day. Our objective is to devise treatment strategies for individuals with AI using a model-based approach.

Several mathematical models of the HPA axis have
been proposed.~\cite{Vadim} developed a model of the HPA axis that accounts for two nonlinear factors: the limit of gland size and the exclusion of non-physiological negative hormone concentrations. In the patient-specific modeling,~\cite{GUDMANDHOEYER201423} identified three potential biomarkers of depression through parameter analysis.~\cite{Karin} examined the long-term stress that leads to hormonal imbalances caused by changes in the mass of corticotrophs, which secrete ACTH and are present in the pituitary gland, and the adrenal cells that secrete cortisol.~\cite{Sriram} identified normal, depressed, and PTSD-affected individuals based on negative cortisol feedback.~\cite{Gupta} developed an HPA axis model emphasizing bistability to characterize stress-related hypocortisolism, thereby providing both short- and long-term physiological responses to stress.~\cite{Markovic} proposed a mathematical model to understand the body's response to acute and chronic stress. The~\cite{ANDERSEN2013122} model focused on stability and bifurcation.~\cite{HOSSEINICHIMEH201552} reviewed five existing models, analyzing anomalies. They selected the model from~\cite{ANDERSEN2013122}. They calibrated it using empirical data from 17 healthy individuals, adjusting the model parameters to reduce the mean absolute percent error by 71\% relative to the available data.~\cite{BANGSGAARD201724} further modified the model of~\cite{ANDERSEN2013122} by incorporating patient-specific parameter estimation, and in addition, modeling circadian and ultradian rhythms. 

In addition to mathematical modeling, various control methodologies have been developed to automate drug dosing in therapeutic applications. A stochastic model-predictive control scheme is proposed to compute drug dosing for cancer therapy in~\cite{HERNANDEZRIVERA2025112255}.~\cite{10644207} proposed a combined MPC and observer-based compensator to generate thyroid replacement therapy for patients with hypothyroidism. Based on the concept of targeted therapy, a model predictive control approach was proposed by~\cite{BenZvi2009} to correct hypocortisolism using simulation-based strategies to improve treatment. Analogously,~\cite{Ankush} presented explicit observer-based nonlinear predictive control to assist in rectifying the HPA axis dysfunctions. They proposed an unknown-input observer to estimate hormone levels and stress from measured ACTH levels.

In our paper, we modify the model of~\cite{ANDERSEN2013122} by incorporating the concept of circadian rhythm from~\cite{BANGSGAARD201724}. We also introduced strong negative feedback from cortisol to CRH to simulate AI condition, as suggested by~\cite{Sriram}. We then use this mathematical model of the HPA axis to generate MPC-based optimal treatment strategies to effectively regulate cortisol levels. Our paper builds upon preliminary results reported in~\cite{Joshi2025}.

\subsection{Paper's Contribution}
Our paper's contributions include:
\begin{enumerate}
    \item Inclusion of circadian and ultradian rhythms in the novel proposed model of the HPA axis.  
    \item Simulation of two conditions of adrenal insufficiency: (a) primary adrenal insufficiency and (b) secondary adrenal insufficiency. 
    \item Introduction of an input variable $u$, to model cortisol replacement for individuals with AI. 
    \item Model verification using published patient data for primary AI.
    \item Construction of a Model Predictive Controller (MPC) using the proposed HPA axis model to generate optimal cortisol replacement therapy for regulating cortisol levels for individuals with AI.
\end{enumerate}

\section{Proposed Model of the HPA Axis}
In this section, we propose a novel HPA axis model for two types of AI: primary adrenal insufficiency, also referred as adrenal disorder, and  secondary adrenal insufficiency, due to a hypothalamic-pituitary disorder. The proposed model integrates the model of~\cite{ANDERSEN2013122}, combines the endogenous circadian rhythm function from~\cite{BANGSGAARD201724}, and incorporates the strong negative feedback from cortisol to CRH as suggested by~\cite{Sriram}. 

 The HPA axis model proposed by~\cite{ANDERSEN2013122} consists of a hippocampal feedback mechanism that takes into account glucocorticoid receptors (GR) and mineralocorticoid receptors (MR). Cortisol binds to GR in body tissues and provides negative feedback to the hypothalamus and pituitary gland. In our proposed model, we incorporate the circadian rhythm function, $C(t)$, from~\cite{BANGSGAARD201724}. The resulting model has the form:
\begin{align}
\frac{dx_1}{dt} &=k_0\left( 1  - \psi \frac{x_3^\gamma}{x_3^\gamma + c_3^\gamma} \right)x_3+\textit{C(t)} - w_1 x_1\label{tag1} \\[10pt]
\frac{dx_2}{dt} &= k_1 \left( 1 - \rho \frac{x_3^\alpha}{x_3^\alpha + c^\alpha} \right)x_1 - w_2 x_2 \label{tag2} \\[10pt]
\frac{dx_3}{dt} &= k_2 x_2 - w_3 x_3+u \label{tag3}\\
C(t) &= N_c \left( \frac{t_m^k}{t_m^k + \alpha_1^k} \cdot \frac{(T - t_m)^l}{(T - t_m)^l + \beta^l} + \epsilon \right), \label{tag4}
\end{align}
where variables $x_1$, $x_2$, and $x_3$, refer to the concentration of CRH, ACTH, and cortisol. The parameters $k_0$, $k_1$, $k_2$ are the positive stimulus rates, and $\omega_1$, $\omega_2$, $\omega_3$ are depletion rates of CRH, ACTH, and cortisol, respectively. The parameter $\alpha$ represents the number of cortisol molecules that bind to free receptors, and $\psi$ is an inhibition constant of CRH due to the binding of cortisol to MR. The parameters $\alpha$, $\gamma$, $c$, and $c_3$ are derived from the chemical stoichiometric equations for receptors, while $\rho$ depends on the receptor count~\cite[page~127]{ANDERSEN2013122}. The cortisol replacement process is modeled as $u$ in~\eqref{tag3}.

The circadian rhythm function is modeled by~\eqref{tag4}, where $t \in [0, T]$, with $T = 1440$ minutes. The function $C(t)$ is bounded in the range $[\epsilon, 1]$, where $\epsilon$ is a small positive design parameter. The value of $t_m$ is calculated as $t_m$ = $(t -\delta) \mod T$, where $\delta$ is the time shift, and $N_c$, a design parameter, is the normalization constant. The parameter $k$ is the steepness of the increasing Hill function at $t_m = \alpha_1$, while $l$ is the steepness of the decreasing Hill function at $t_m = \beta$, where $\alpha_1$ and $\beta$ are two modeling constants. The values of the above parameters for circadian rhythm function and the two AI conditions are presented in Table~\ref{tab:parameters1} and Table~\ref{tab:parameters2}.

\begin{table}[h!]
\centering
\caption{Parameters of circadian rhythm function $C(t)$ given by~\cite{BANGSGAARD201724}.}
\begin{tabular}{@{}clll@{}}
\hline
\textbf{No.} & \textbf{Parameter} & \textbf{Value} & \textbf{Unit} \\ 
\hline
1 & $\delta$  & $83.8$                & min \\ 
2 & $\alpha_1$  & $300$               & min \\ 
3 & $\beta$   & $950$                 & min \\ 
4 & $k$       & $5$                   & -- \\ 
5 & $l$       & $6$                   & -- \\ 
6 & $\epsilon$ & $0.01$               & -- \\ 
7 & $N_c$     & $0.5217$              & -- \\ 
\hline
\end{tabular}
\label{tab:parameters1}
\end{table}

\begin{table}[h!]
\centering
\caption{Parameters used to model an individual with adrenal insufficiency.}
\label{tab:parameters2}
\begin{tabular}{@{}cllll@{}} 
\hline
\textbf{No.} & \textbf{Parameter} & \textbf{Primary AI} & \textbf{Secondary AI} & \textbf{Unit} \\ 
\hline
1  & $k_0$    &43.809    & 0.12885 & pg/(mL$\times$min) \\ 
2  & $k_1$    & 0.02  & 0.0055 & min$^{-1}$ \\ 
3  & $k_2$    & 0.00001  &0.0009 & min$^{-1}$ \\ 
4  & $w_1$    & 0.12731  & 0.12731 & min$^{-1}$ \\ 
5  & $w_2$    & 0.0348  & 0.0348 & min$^{-1}$ \\ 
6  & $w_3$    & 0.001871 & 0.001871 & min$^{-1}$ \\ 
7  & $\rho$   & 0.5     & 0.5 & -- \\ 
8  & $\psi$   & 1     & 1 & -- \\  
9 & $\alpha$ & 1       & 1 & -- \\ 
10 & $\gamma$ & 5       & 5 & -- \\ 
11 & $c$      & 5.06    & 5.06 & pg/mL \\ 
12 & $c_3$    & 1.42    & 1.42 & pg/mL \\ 
\hline
\end{tabular}
\end{table}

\begin{figure}
\centerline{\includegraphics[width=1.75\textwidth]{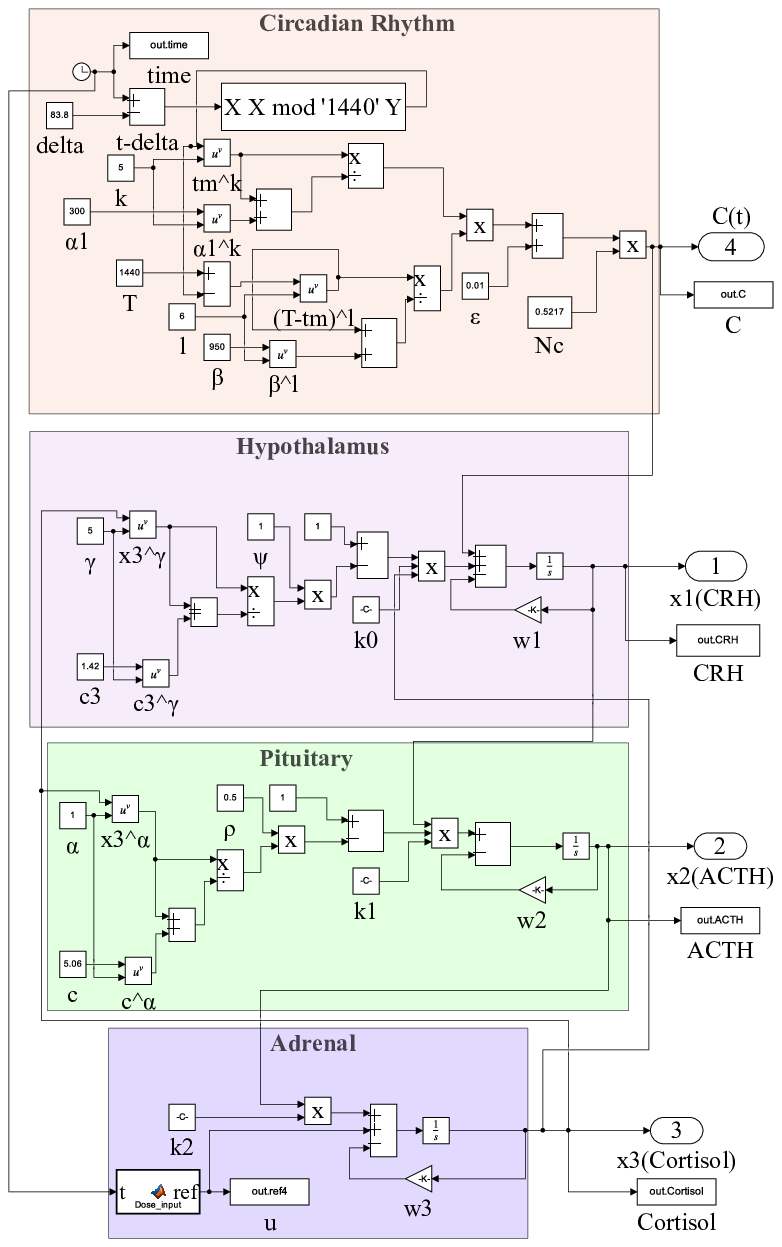}}
\caption{Simulink diagram of the proposed HPA axis model.}
\label{fig:Simulink}
\end{figure}

Fig.~\ref{fig:Simulink} represents a Simulink model of the HPA axis, providing a conceptual understanding of the dynamics of the HPA axis. The detailed Simulink model of the HPA axis is divided into four sections: circadian rhythm, hypothalamus, pituitary, and adrenal glands. Each section consists of interconnected mathematical blocks that represent the differential equations~\eqref{tag4},~\eqref{tag1},~\eqref{tag2},~\eqref{tag3} that control hormonal interactions.

To validate the accuracy of the proposed model, we use an open-loop cortisol replacement medication strategy, which is presently a standard patient treatment. The validation process of the model is discussed in the following section.

\section{Standard Patient Treatment Simulation}
In this section, we describe the cortisol replacement strategy for individuals with AI. As recommended by the Endocrine Society, see~\cite{Bornstein}, cortisol replacement therapy typically involves taking 15--25~mg of hydrocortisone (HC) in two or three divided doses throughout the day. HC is commonly available in 10~mg tablets, which can be divided into quarter doses to tailor the medication to individual patient needs. Typically, the most significant portion is administered in the morning, while smaller doses are given in the afternoon and evening to simulate the diurnal variation.

In our simulation, the first dose accounts for 60\% of the total daily dose, the second and third doses each account for 20\%. For a daily HC dosage of 20~mg, the doses are divided as 12~mg at 8:00~am, 4~mg at 2:00~pm, and 4~mg at 8:00~pm. For a 25~mg daily dosage, the schedule consists of 15~mg in the morning at 8:00~am, 5~mg in the afternoon at 2:00~pm, and 5~mg in the evening at 8:00~pm.

\begin{figure}[H]
\centerline{\includegraphics[width=0.6\textwidth]{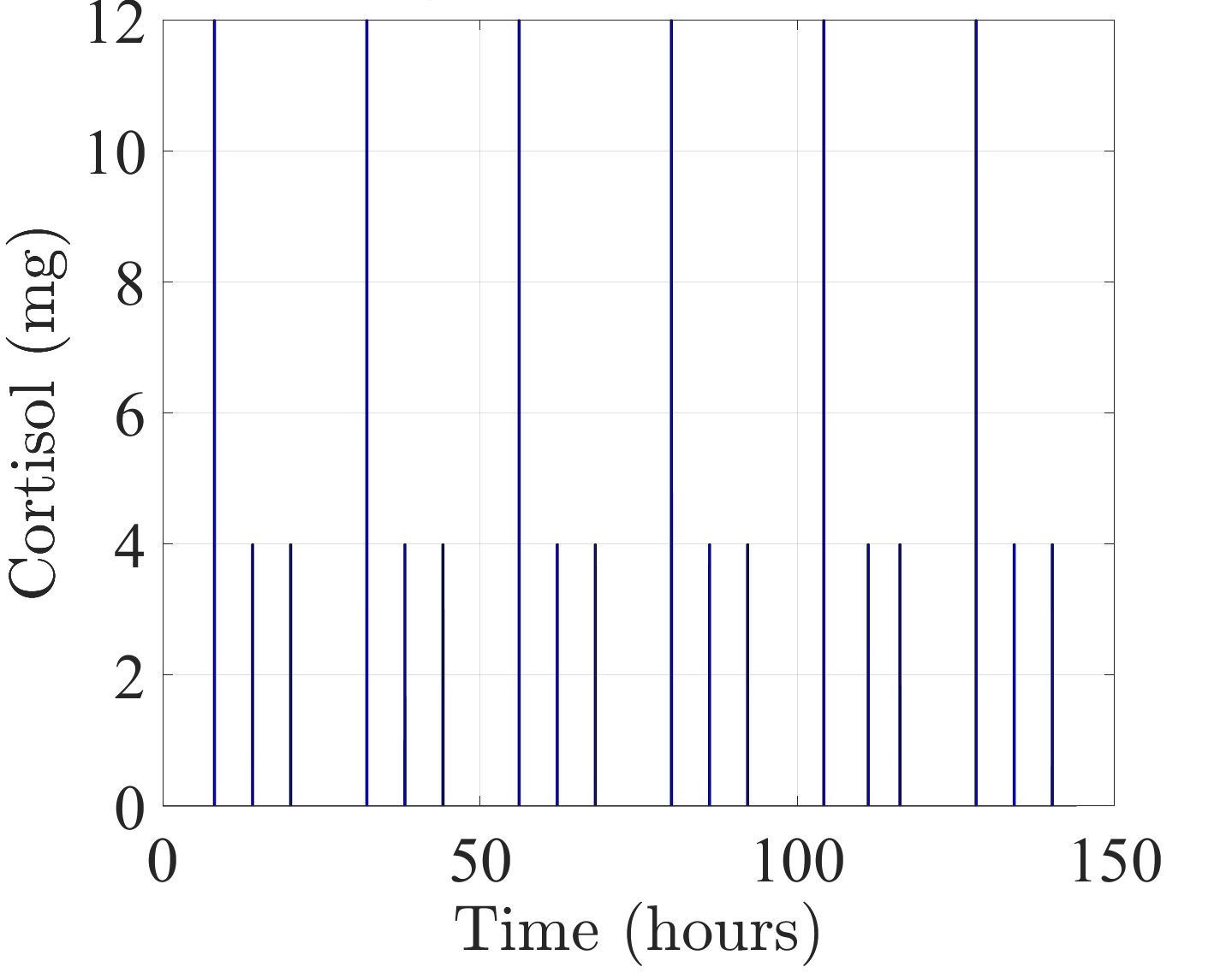}}
\caption{Cortisol replacement therapy with a total daily dose of 20~mg hydrocortisone (HC). Administration of cortisol at 8:00~am, 2:00~pm, and 8:00~pm over six days.}
\label{fig:T1}
\end{figure}

\begin{figure}[H]
\centerline{\includegraphics[width=0.6\textwidth]{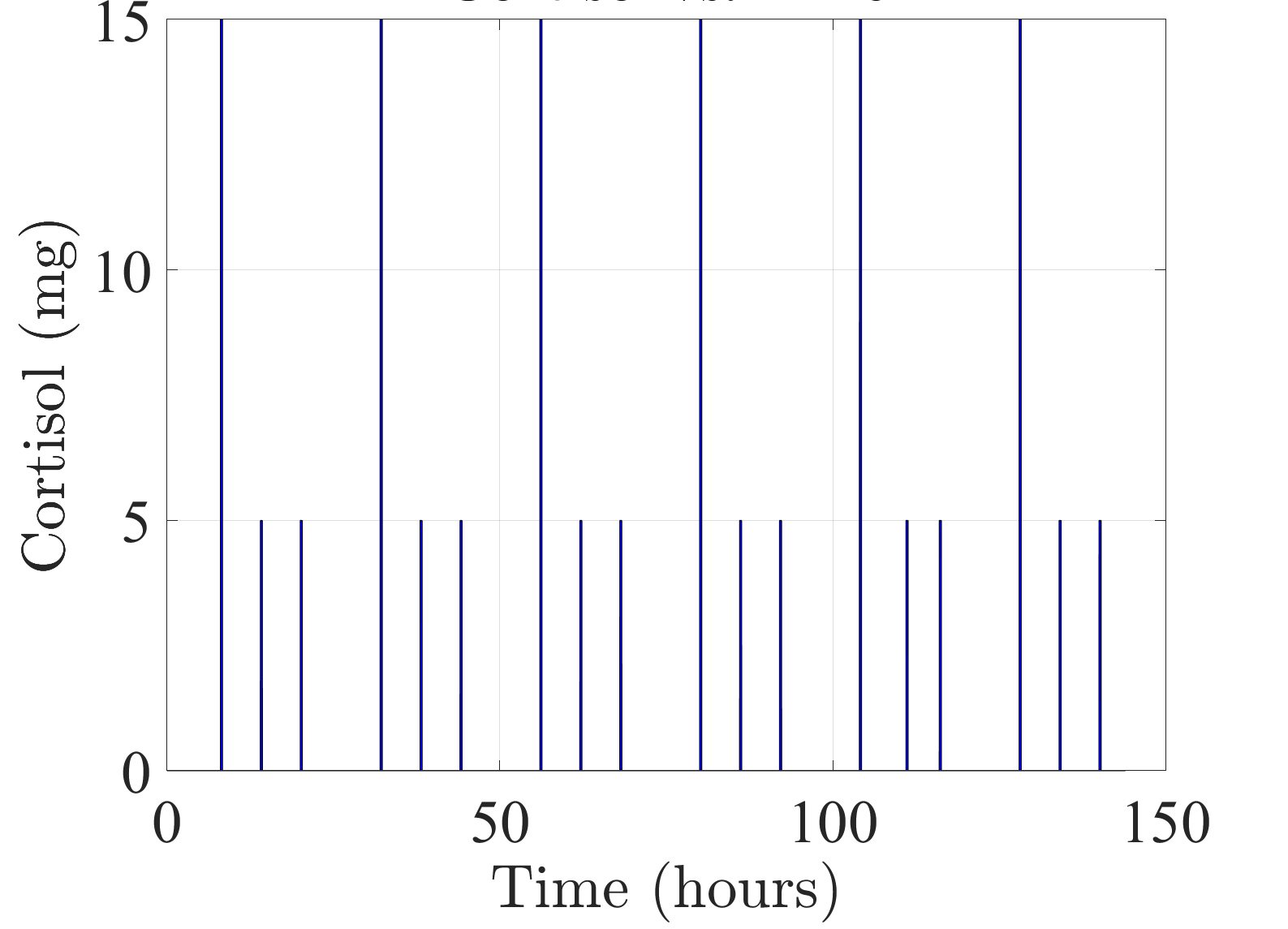}}
\caption{Cortisol replacement therapy with a total daily dose of 25~mg hydrocortisone (HC). Administration of cortisol at 8:00~am, 2:00~pm, and 8:00~pm over six days.}
\label{fig:T2}
\end{figure}

\begin{figure}[H]
\centerline{\includegraphics[width=\textwidth]{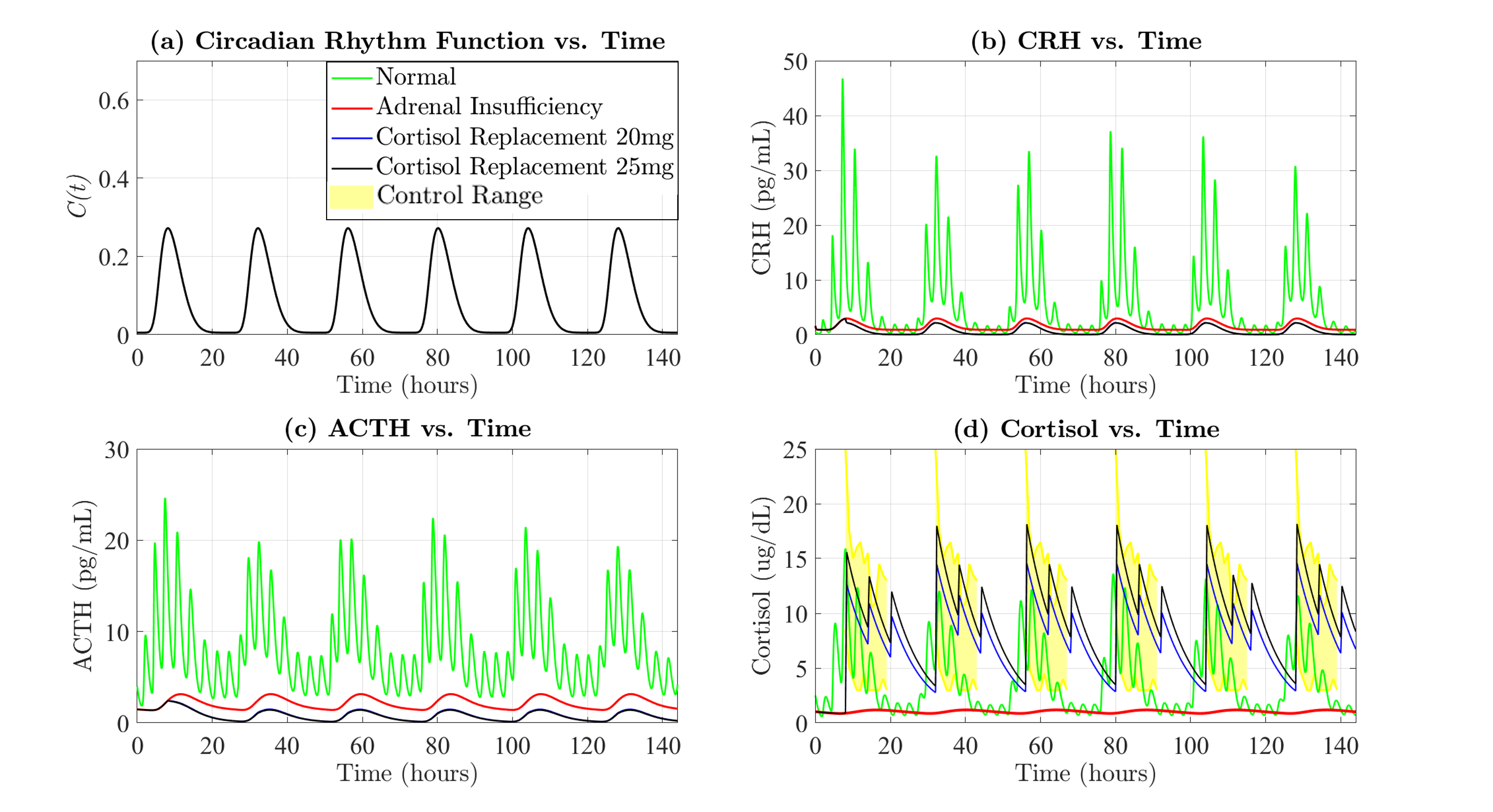}}
\caption{Comparison of normal case, secondary adrenal case, and with cortisol replacement therapy of 20~mg and 25~mg for six days. Plots of levels of CRH, ACTH, and cortisol compared against the circadian rhythm $C(t)$. In all sub-figures, the legends are the same as in sub-figure (a).}
\label{fig:Graph-HP-case}
\end{figure}

We have modeled two conditions of AI using a Simulink model. We compared the hormone levels of a normal individual taken from~\cite{BANGSGAARD201724}, with a secondary AI condition, and with cortisol replacement therapy of 20~mg and 25~mg for six days. We have simulated this scenario using the dosage shown in Fig.~\ref{fig:T1} and Fig.~\ref{fig:T2}, and the simulation results are presented in Fig.~\ref{fig:Graph-HP-case}. From the plots in Fig.~\ref{fig:Graph-HP-case}, we can infer that ACTH and CRH concentrations remain low; therefore, cortisol levels are also low. After administering cortisol replacement therapy for six days, cortisol spikes are observed. The cortisol level increases to a normal range as shown in Fig.~\ref{fig:Graph-HP-case}{(d)}, whereas ACTH and CRH remain low as in Fig.~\ref{fig:Graph-HP-case}{(c)} and Fig.~\ref{fig:Graph-HP-case}{(b)} respectively. Cortisol spikes are lower on the first day than on the subsequent five days, reflecting the gradual accumulation of cortisol following the initial dose of medication. A 25~mg HC dose was observed to produce a more substantial inhibitory effect on CRH and ACTH than a 20~mg medication dose.

\begin{figure}[H]
\centerline{\includegraphics[width=\textwidth]{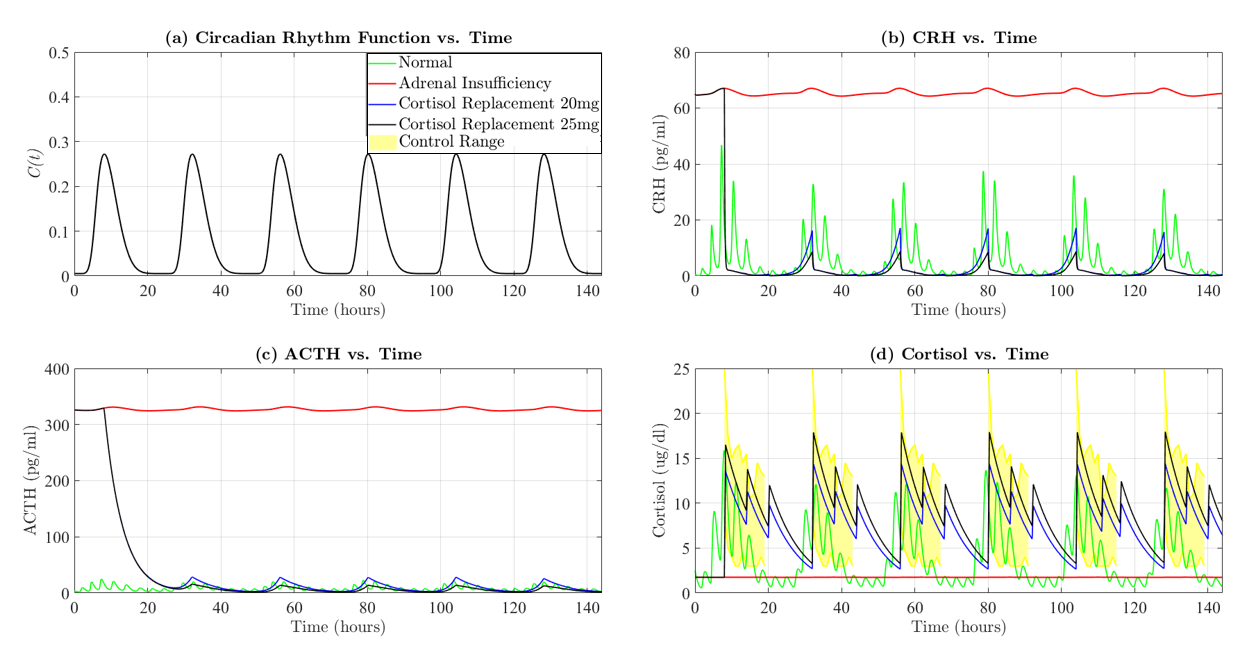}}
\caption{Comparison of normal individual, primary adrenal insufficiency, and cortisol replacement therapy of 20~mg and 25~mg for six days. Plots of levels of CRH, ACTH, and cortisol compared against the circadian rhythm $C(t)$. The legends are the same as in sub-figure (a).}
\label{fig:Adrenal3}
\end{figure}

In Fig.~\ref{fig:Adrenal3}, we show simulation results of the primary AI condition, where we compare the normal case with the adrenal case, with and without medication. Here, we again apply 20~mg and 25~mg of cortisol replacement therapy. After medication, cortisol levels increase and reach the normal range, as shown in Fig.~\ref{fig:Adrenal3}{(d)}. The ACTH concentration drops down suddenly to a certain value that is in the normal range, as in Fig.~\ref{fig:Adrenal3}{(c)}, and CRH concentration also degrades due to the feedback loop via cortisol, as shown in Fig.~\ref{fig:Adrenal3}{(b)}. The 25~mg HC dose exerts stronger negative feedback on CRH and ACTH than the 20~mg dose. As a result, CRH and ACTH levels decrease more rapidly and reach lower magnitudes after the 25~mg administration of cortisol. In this case as well, cortisol spikes are relatively lower on the first day than on the following five days, as the hormone gradually builds up in concentration after the initial medication. 

These simulations are consistent with real patient data from~\cite{Rousseau2015}. The yellow shaded region in Fig.~\ref{fig:Graph-HP-case}{(d)} and Fig.~\ref{fig:Adrenal3}{(d)} represents the control range of plasma cortisol, as given in~\cite{Rousseau2015}. While patients in their study received HC doses distributed as 50\%, 25\%, and 25\% across the day, the proposed model achieves the desired cortisol levels by administering doses of medication in a 60\%, 20\%, and 20\% distribution. This adjustment is based on the proposed mathematical modeling of the HPA axis. 

The simulation results show that it is analytically possible to arrive at a fixed working cortisol replacement strategy. However, this strategy, though effective, is not optimal. To achieve optimal cortisol replacement strategies, a model predictive controller (MPC) is proposed in the following section. 

\section{Constructing a Physician's Virtual Assistant Using Model Predictive Control (MPC)}
Model Predictive Control (MPC) is a control method that makes use of a process model to generate control strategies that satisfy constraints on the control inputs. The model of the process is used to generate the sequence of predicted outputs. This sequence of predicted outputs is compared against the desired sequence of outputs. An optimization process is used to minimize the error between the desired sequence of outputs and the predicted sequence of outputs. The MPC solves an optimization problem using the current state and applies only the first portion of the control sequence at the current time. The computational process is then repeated by taking a new state measurement and computing a new optimal control sequence. This approach is also known as moving-horizon control or receding-horizon control; for more information on MPC see~\cite{Maciejowski2002PredictiveC}.

In a clinical study by~\cite{hammarstrand2017glucocorticoid}, it was found that higher cortisol replacement therapy was associated with increased mortality.
Both overdosing and underdosing can be detrimental to one's health. Therefore, we propose an MPC-based controller to generate varying doses of cortisol replacement therapy for individuals with AI. The aim is to regulate the cortisol hormone concentrations of an individual with AI toward a normal level while adhering to constraints on the amount of medication. These constraints on the medication are enforced for two main reasons: (i) medication must be non-negative, and (ii) the medication prescribed cannot exceed a specific amount, as too much cortisol could have adverse effects on the patient.

\subsection{Model Linearization and Discretization}
The proposed model of the HPA axis given by~\eqref{tag1}, \eqref{tag2}, \eqref{tag3}, and \eqref{tag4} is highly non-linear and needs to be linearized to design an MPC controller. This linearized model, after discretization, is referred to as the design model. We linearize the non-linear model about an equilibrium state corresponding to $u=0$. The equilibrium state is computed by solving a set of algebraic equations obtained by setting the state derivatives to zero. The components of the equilibrium state vector are CRH, ACTH and cortisol concentrations, denoted $x_{1eq}$, $x_{2eq}$, and $x_{3eq}$. There numerical values are obtained using the MATLAB's function \texttt{fsolve}. The circadian rhythm function is averaged to obtain a steady-state solution. The equilibrium state for both adrenal insufficiency conditions is obtained using the parameters given in Table~\ref{tab:parameters2}. We construct the Jacobian matrix and substitute the equilibrium state vector to get the linearized HPA axis model. The continuous time linearized model for $u=0$ has the form: 
\begin{equation}
\delta\dot{x}(t) = J(x)\rvert_{x = x_{\text{eq}}} \, \delta x(t).
\tag{5}
\end{equation}
The perturbation state $\delta x(t) =x(t) -x_{eq}$ represents deviation from the equilibrium (see~\cite{zak2003systems} for more on linearization). Once the nonlinear system is linearized, it is discretized using an exact discretization method using a zero-order hold (ZOH). We obtain the discretized system using the MATLAB command~\texttt{c2d}. The sampling time used is $T_s = 1$~hour. The discretized model has the form,
\begin{equation}
x[k+1] = A_dx[k].
\tag{6} 
\end{equation}

\subsection{Virtual Assistant for Cortisol Regulation}
\begin{figure}[H]
\centerline{\includegraphics[width=\textwidth]{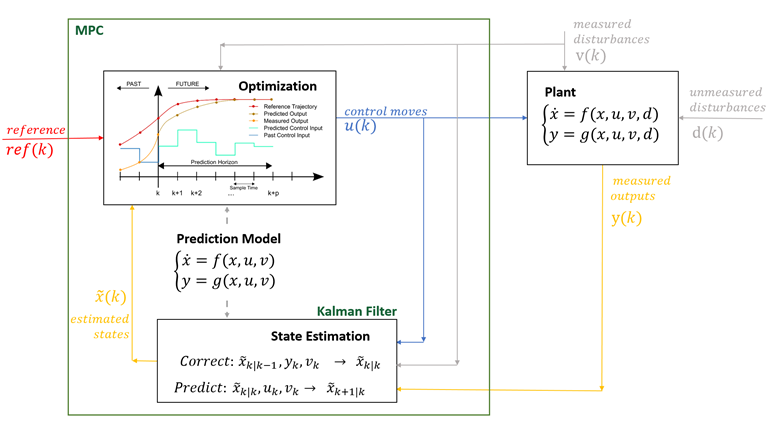}}
\caption{Block diagram of MPC~\cite{mathworksMPC}.}
\label{fig:mpc}
\end{figure}

In this section, we use MATLAB's Model Predictive Control Toolbox to construct a physician virtual assistant to regulate cortisol levels in individuals with AI. The linearized discrete model discussed previously is used to construct our virtual assistant, as illustrated in Fig.~\ref{fig:mpc}. The controller receives the nonlinear patient model output and uses a linearized model to estimate the patient's state and calculate a sequence of control moves that minimize a cost function over a given prediction horizon. 

We configure the MPC controller using MATLAB's built-in~\texttt{mpc()} function, with a prediction horizon of $N_p=10$ and a control horizon of $N_c=4$. The prediction horizon shows how far the controller "looks into the future." The control horizon is the parameter for which the controller computes $N_c$ free control samples occurring at times $k$ through $k+N_c-1$, and holds the controller output constant for the remaining prediction horizon steps from $k+N_c$ through $k+N_p-1$. Here, $k$ is the discrete time. The sampling time is $T_s=1$~hour. Recall that the state-space model of the plant includes three states: CRH, ACTH, and cortisol concentrations. The controlled input affects the rate of change of cortisol concentrations. Cortisol concentration is the only measured output, while CRH and ACTH are unmeasured states. 

The reference signal $r[\cdot]$ represents the target cortisol trajectory, and the error between the reference and the predicted output is calculated using a Quadratic Programming (QP) solver that penalizes deviations from the reference. The optimal control signal $u[k]$ is subject to input constraints of $u_{\text{min}} = 0$~mg and $u_{\text{max}} = 15$~mg per dose. Also, a daily limit of $30$~mg cortisol replacement therapy is imposed to prevent overdosing. In clinical trials, the MPC controller's prescribed input doses should align with the medical doses clinically available. It is impractical to administer 5.45~mg, even though such a dose may theoretically yield the best results for tracking desired reference trajectories. Thus, the medication is quantized, and cortisol doses are rounded to the nearest 5~mg dose. The dosing times are set to 8:00~am, 2:00~pm, and 8:00~pm, to align with the clinical practices.

At each time step, the controller receives the measured cortisol value and calculates the optimal cortisol replacement therapy to regulate cortisol levels toward a reference value. Internally, MATLAB's MPC uses a default Kalman filter to estimate the patient state from the measured output, which is then used to predict future behavior and solve the optimization problem as illustrated in Fig.~\ref{fig:mpc}. The predicted input is then applied to the non-linear patient model, which is simulated using~\texttt{ode45} over the next six hours to obtain the value of the states. The whole process is repeated for six days. Simulations confirm that the MPC can effectively regulate cortisol within the desired range. This demonstrates the potential of an MPC-based virtual assistant in guiding hormone replacement therapy. In the following subsections, we discuss the application of the MPC-based virtual assistant to obtain a cortisol replacement strategy for both primary and secondary AI conditions. 

\subsubsection{Primary Adrenal Insufficiency}
\begin{figure}
\centerline{\includegraphics[width=0.65\textwidth]{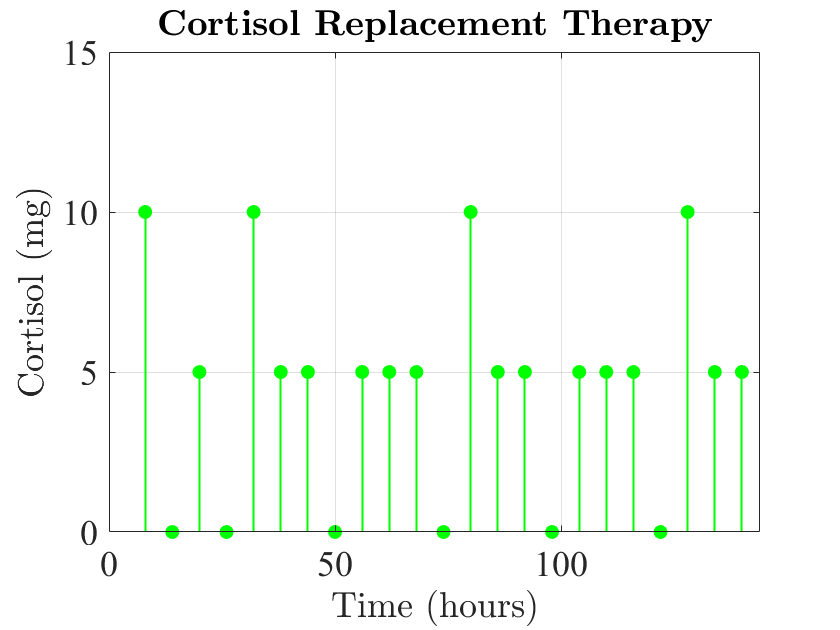}}
\caption{Input generated by MPC to regulate cortisol concentration around 10~$\mu$g/dL for primary AI.}
\label{fig:MPC10A}
\end{figure}

\begin{figure}[H]
\centerline{\includegraphics[width=\textwidth]{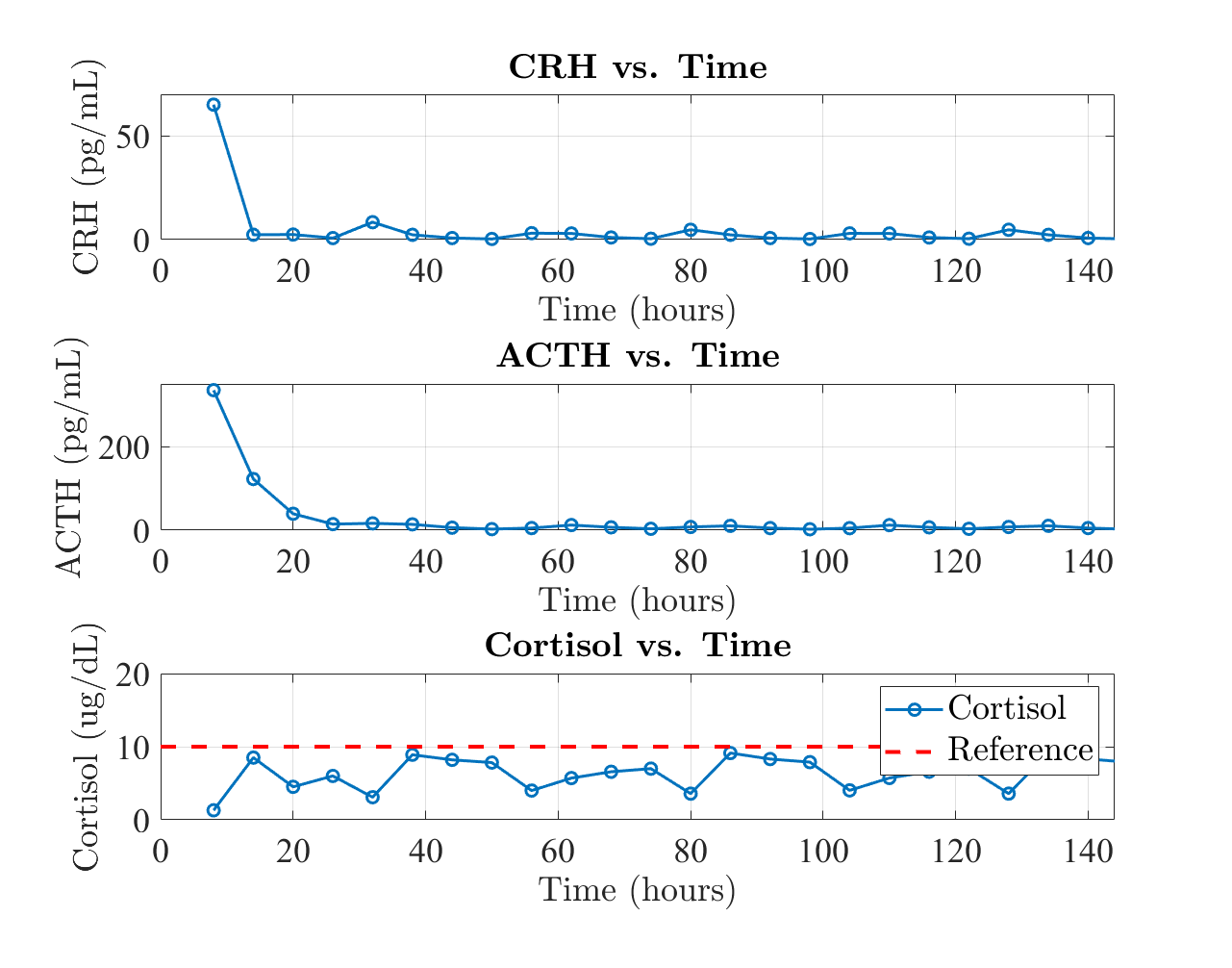}}
\caption{Trajectories of CRH, ACTH, and cortisol concentration in response to the MPC-controlled input for primary adrenal insufficiency. The controller regulates cortisol levels to track a reference concentration of 10~$\mu$g/dL.}
\label{fig:MPC10AP}
\end{figure}

In this subsection, we evaluate the effectiveness of an MPC-based virtual assistant to control primary AI using various cortisol reference trajectories. Fig.~\ref{fig:MPC10A} shows the cortisol replacement therapy input generated by the MPC controller for a reference concentration of 10~$\mu$g/dL. Fig.~\ref{fig:MPC10AP} depicts the state trajectories of CRH, ACTH, and cortisol after the administration of controlled input from MPC. The controller forces the cortisol to track the reference cortisol profile accurately. The levels of CRH and ACTH are elevated in the absence of cortisol but decline following administration of MPC-prescribed doses, consistent with the expected negative feedback.

\begin{figure}
\centerline{\includegraphics[width=0.65\textwidth]{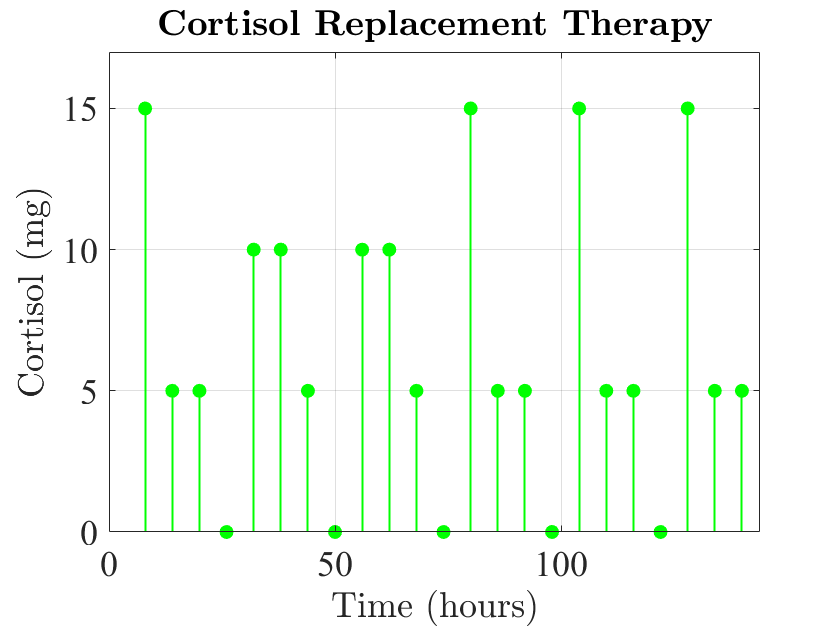}}
\caption{Input generated by MPC to regulate cortisol concentration around 15~$\mu$g/dL for primary AI.}
\label{fig:MPC15A}
\end{figure}

\begin{figure}[H]
\centerline{\includegraphics[width=\textwidth]{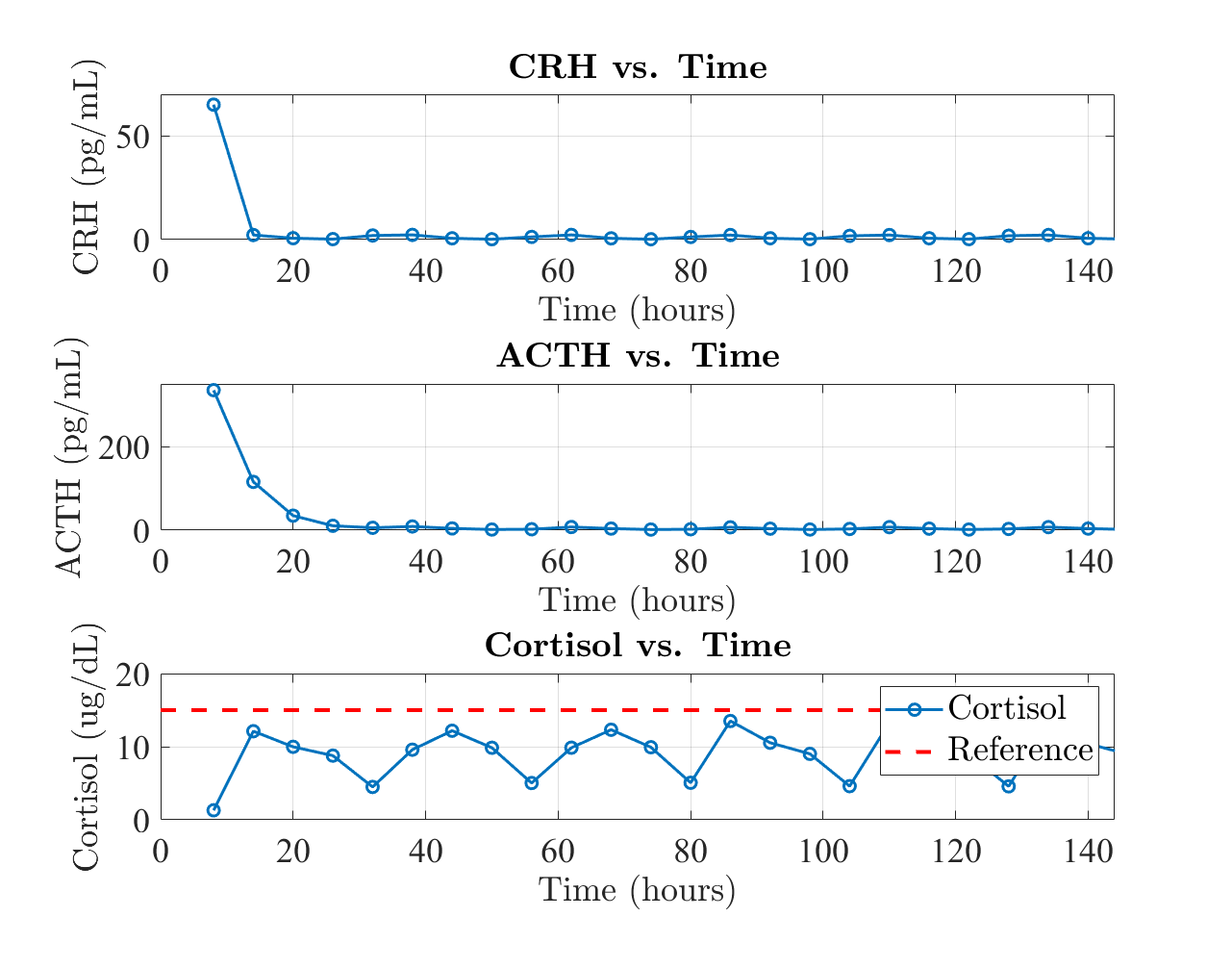}}
\caption{Trajectories of CRH, ACTH, and cortisol concentration in response to the MPC-controlled input for primary adrenal insufficiency. The controller regulates cortisol levels to track a reference concentration of 15~$\mu$g/dL.}
 \label{fig:MPC15AP}
\end{figure}

Fig.~\ref{fig:MPC15A} shows the MPC-controlled cortisol replacement therapy input for a reference cortisol concentration trajectory of 15~$\mu$g/dL. Fig.~\ref{fig:MPC15AP} shows the CRH, ACTH, and cortisol concentrations after medication. As in the previous case, the hormone trajectories demonstrate effective tracking of the reference, with suppressed CRH and ACTH levels after medication administration.

\begin{figure}
\centerline{\includegraphics[width=0.65\textwidth]{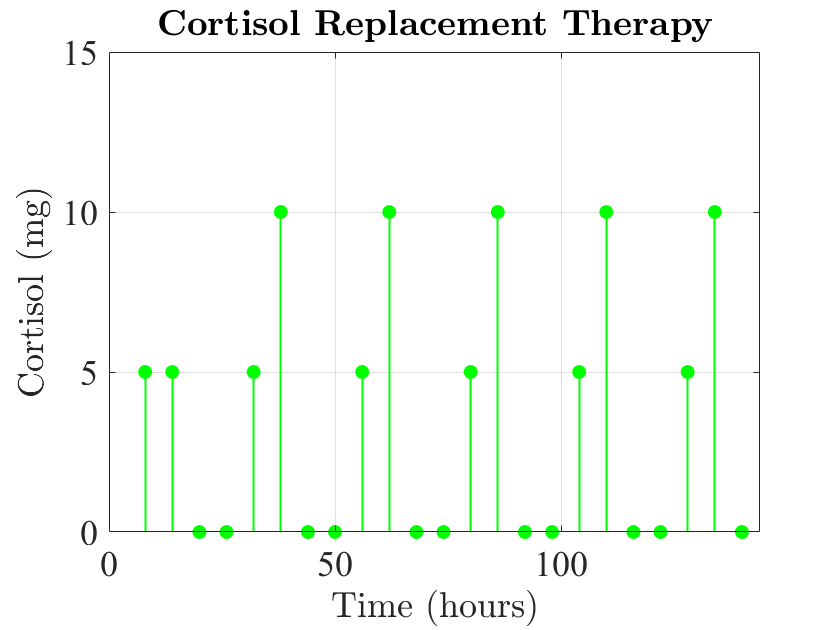}}
\caption{Input generated by MPC to regulate cortisol concentration around a sinusoidal function for primary AI.}
\label{fig:MPCsinA}
\end{figure}

\begin{figure}[H]
\centerline{\includegraphics[width=\textwidth]{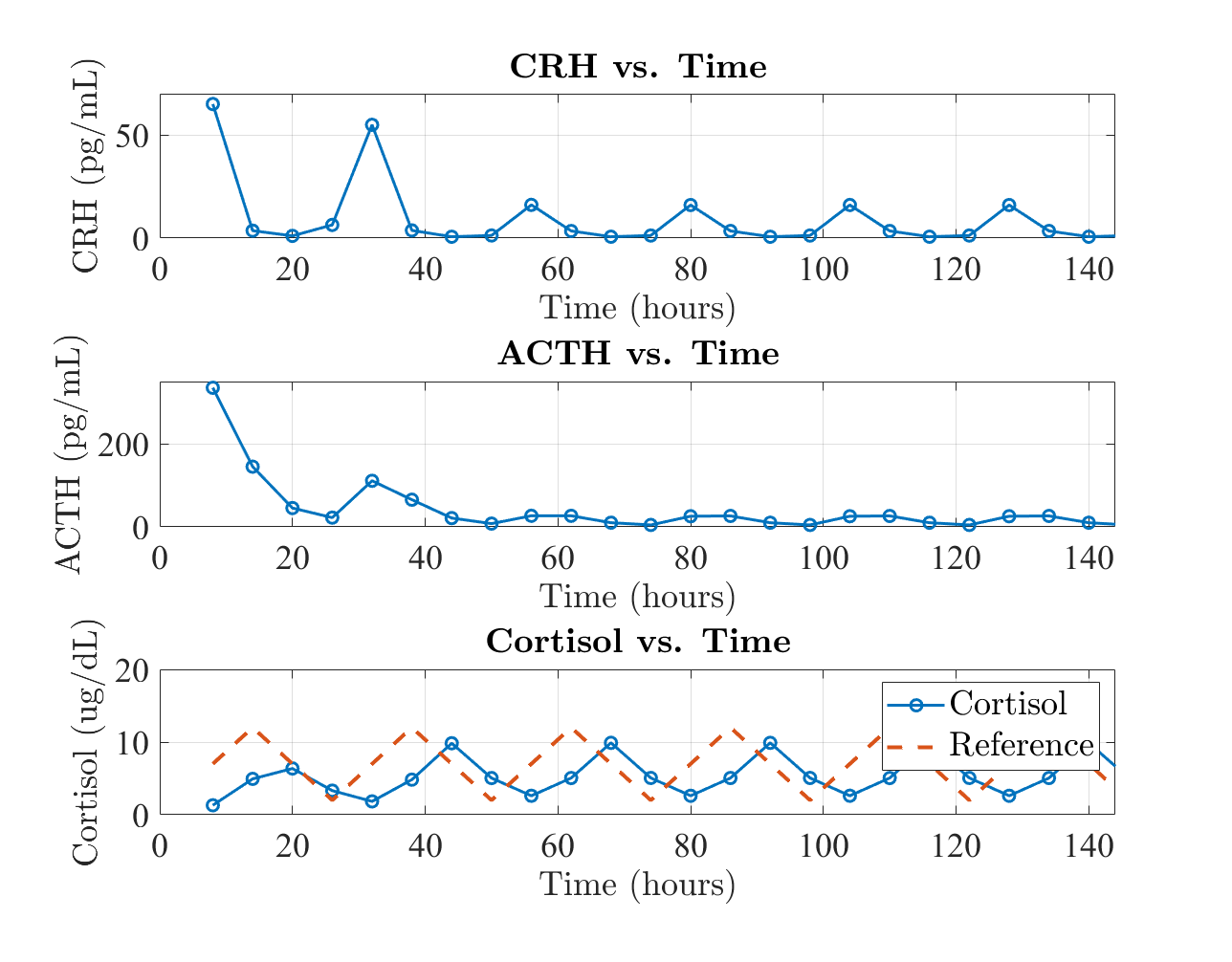}}
\caption{Trajectories of CRH, ACTH, and cortisol concentration in response to the MPC-controlled input for primary adrenal insufficiency. The controller regulates cortisol levels to track a reference concentration of sinusoidal function.}
\label{fig:MPCsinAP}
\end{figure}

Finally, a sinusoidal reference trajectory is selected to mimic a changing reference of the cortisol profile throughout the day. The MPC-generated inputs are as shown in Fig.~\ref{fig:MPCsinA}. The corresponding hormone responses are shown in Fig.~\ref{fig:MPCsinAP}. The controller can adapt the dosage pattern to follow the changing cortisol demand. A slight phase lag is observed between the administered cortisol and the reference. However, from the physiological perspective, the administered cortisol trajectory is appropriate.

\subsubsection{Secondary Adrenal Insufficiency}
\begin{figure}[H]
\centerline{\includegraphics[width=0.65\textwidth]{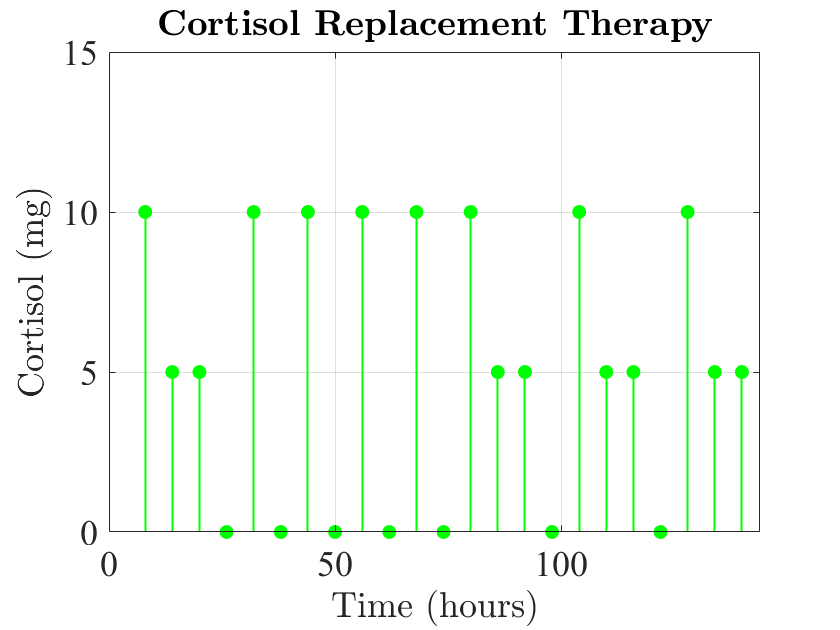}}
\caption{Input generated by MPC to regulate the cortisol concentration around 10~$\mu$g/dL for secondary AI.} 
\label{fig:MPC10H}
\end{figure}

\begin{figure}[H]
\centerline{\includegraphics[width=\textwidth]{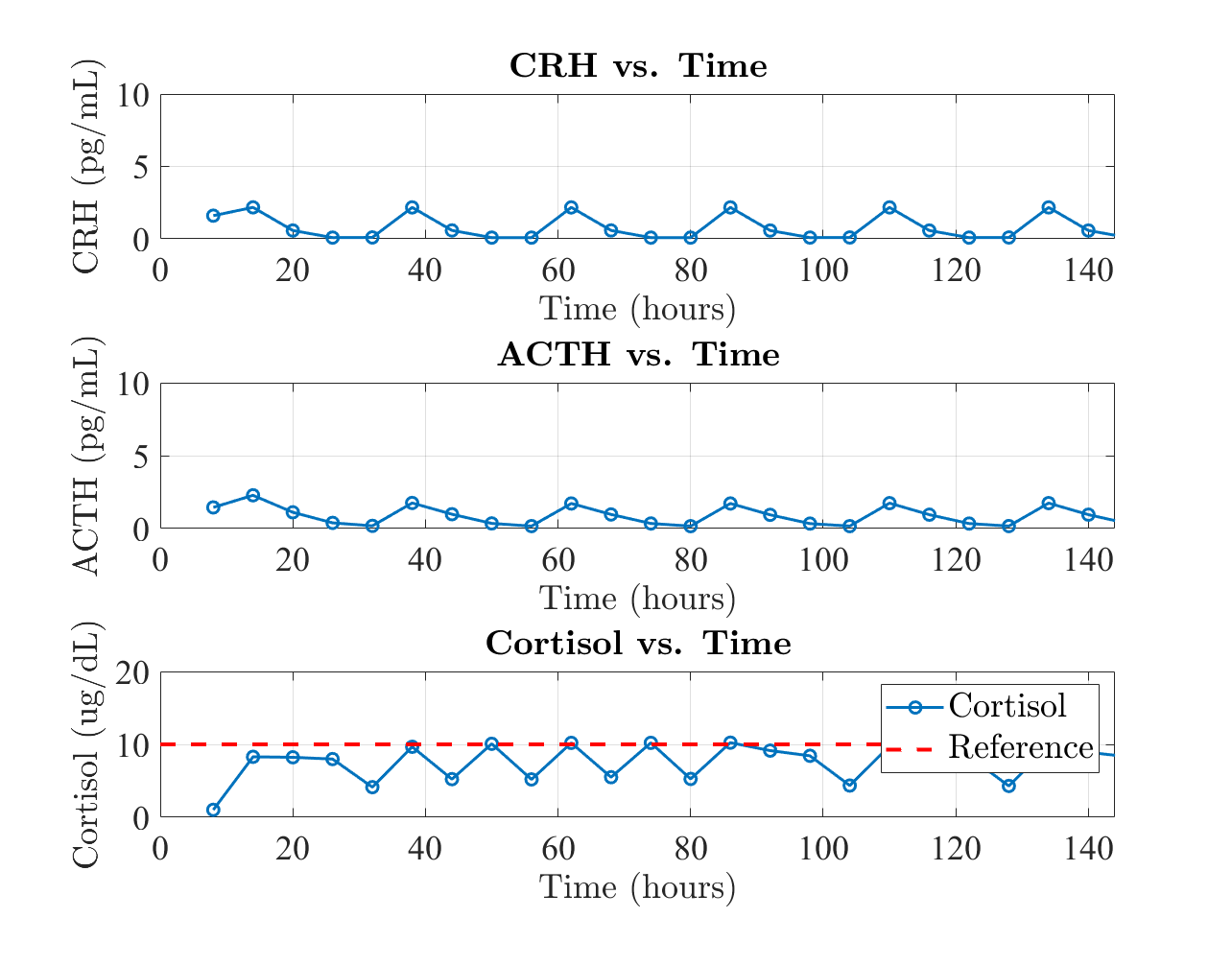}}
\caption{Trajectories of CRH, ACTH, and cortisol concentration in response to the MPC-controlled input for secondary adrenal insufficiency. The controller regulates cortisol levels to track a reference concentration of 10~$\mu$g/dL.}
\label{fig:MPC10HP}
\end{figure}

As in the primary AI, we select various reference trajectories for the cortisol hormone to simulate treatment strategies, thereby validating an MPC-based virtual assistant developed for individuals with secondary AI. Fig.~\ref {fig:MPC10H} depicts the cortisol replacement therapy generated by the MPC controller to regulate the cortisol concentration around a reference of 10~$\mu$g/dL. Fig.~\ref {fig:MPC10HP} shows the trajectories of CRH, ACTH, and cortisol concentrations in response to the MPC-controlled input. The controller accurately drives cortisol to the desired reference trajectory. The concentrations of CRH and ACTH remain suppressed even after medication administration, consistent with observations in the open-loop simulations for secondary AI shown in Fig.~\ref{fig:Graph-HP-case}. 

\begin{figure}
\centerline{\includegraphics[width=0.65\textwidth]{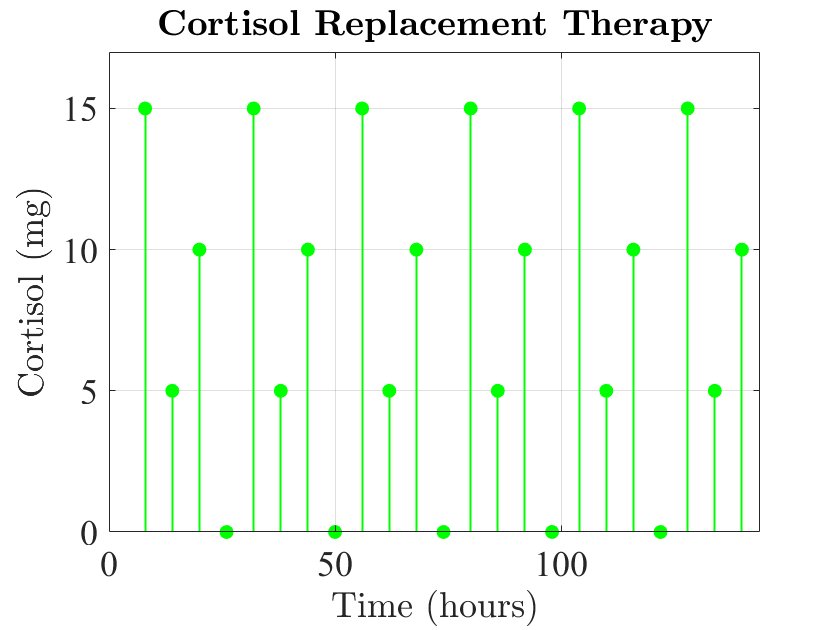}}
\caption{Input generated by MPC to regulate cortisol concentration around 15~$\mu$g/dL for secondary AI.}
\label{fig:MPC15H}
\end{figure}

\begin{figure}[H]
\centerline{\includegraphics[width=\textwidth]{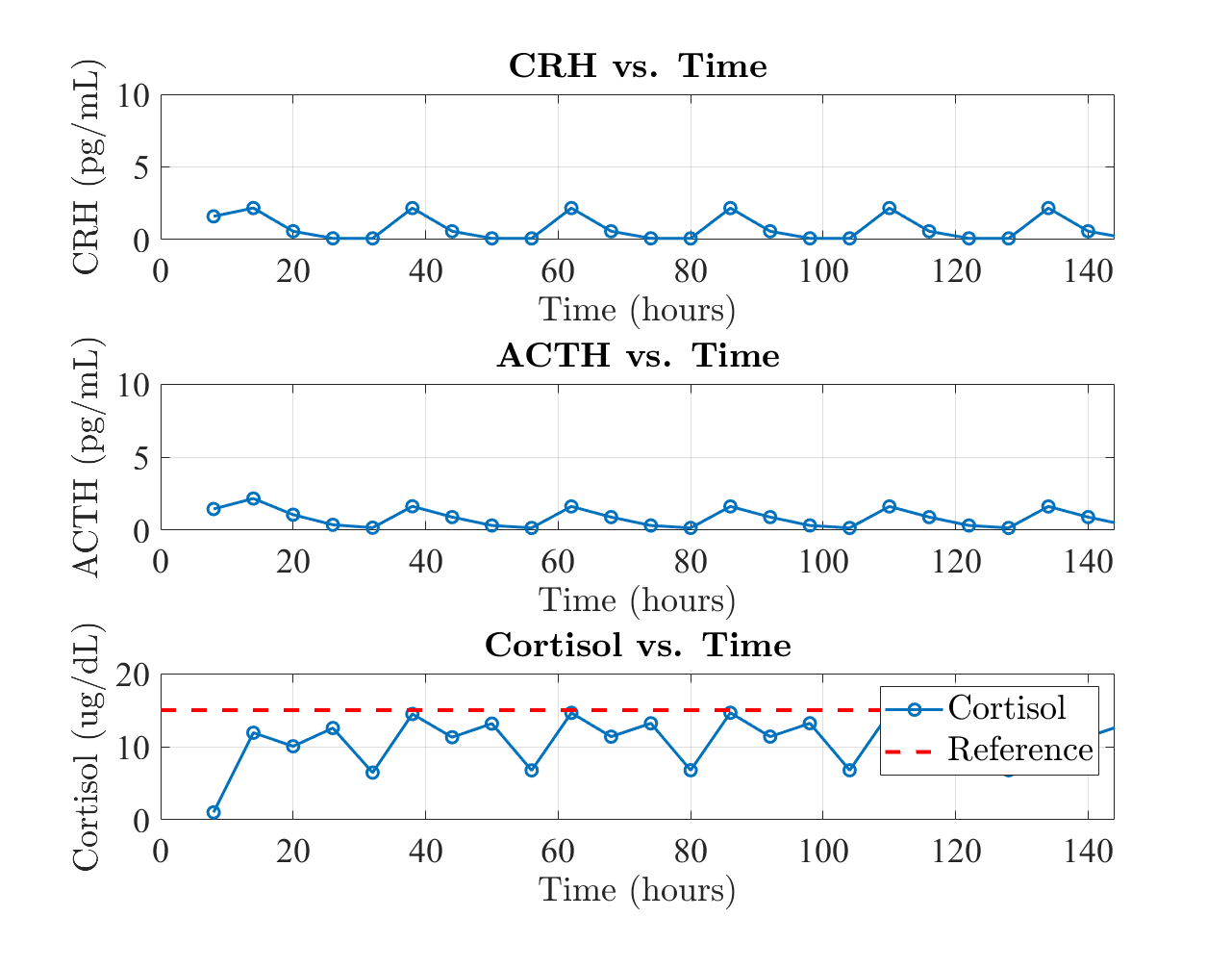}}
\caption{Trajectories of CRH, ACTH, and cortisol concentration in response to the MPC-controlled input for secondary adrenal insufficiency. The controller regulates cortisol levels to track a reference concentration of 15~$\mu$g/dL.}
\label{fig:MPC15HP}
\end{figure}

Fig.~\ref{fig:MPC15H}, shows the inputs generated by MPC to regulate the cortisol concentration around 15~$\mu$g/dL. Fig.~\ref{fig:MPC15HP} shows the state trajectories of CRH, ACTH, and cortisol after medication. The corresponding hormonal response indicates that CRH and ACTH levels remain low throughout the simulation, confirming the expected negative feedback following the administration of the cortisol replacement therapy.  

\begin{figure}
\centerline{\includegraphics[width=0.65\textwidth]{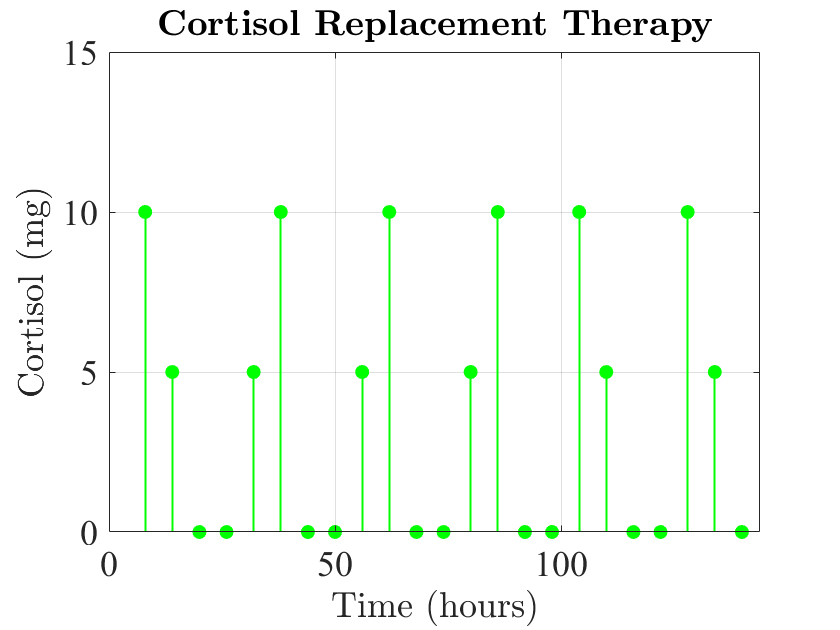}}
\caption{Input generated by MPC to regulate cortisol concentration around a sinusoidal function for secondary AI.}
\label{fig:MPCsinH}
\end{figure}

\begin{figure}[H]
\centerline{\includegraphics[width=\textwidth]{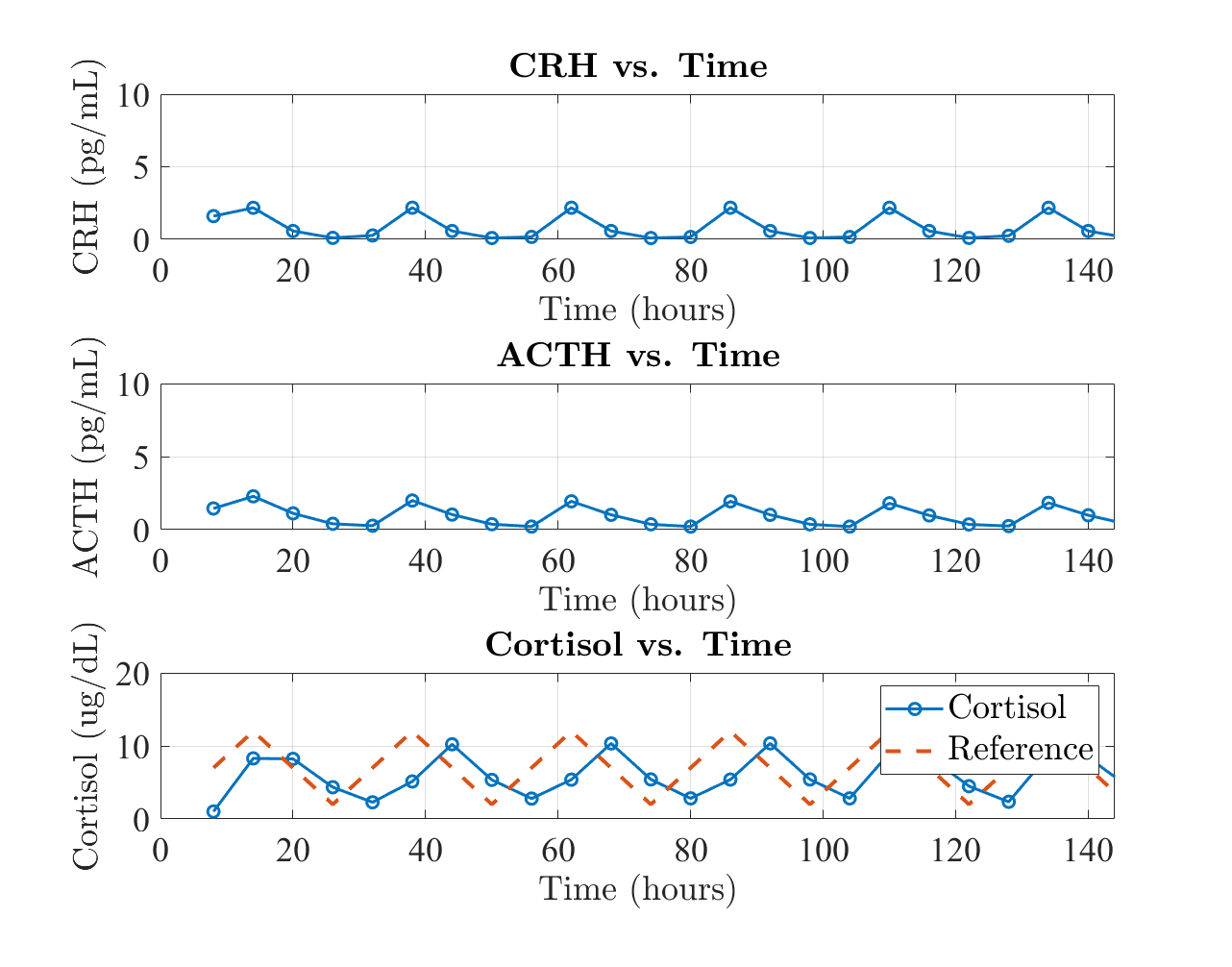}}
\caption{Trajectories of CRH, ACTH, and cortisol concentration in response to the MPC-controlled input for secondary adrenal insufficiency. The controller regulates cortisol levels to track a reference concentration of sinusoidal function.}
\label{fig:MPCsinHP}
\end{figure}

Fig.~\ref{fig:MPCsinH} depicts the  MPC-generated input, and Fig.~\ref{fig:MPCsinHP} shows the state trajectories of CRH, ACTH, and cortisol in response to the cortisol replacement therapy. The reference trajectory in this case is similar to the cortisol profile observed throughout the day, with low levels in the morning and evening and high levels in the afternoon. The MPC successfully adjusts the dosing to follow the target. Cortisol is slightly out of phase with the reference trajectory; however, it follows the physiology very closely.

Our simulation results demonstrate that the MPC-based virtual assistant generates cortisol replacement strategies yielding serum cortisol at desired levels.

\section{Conclusion and Future Scope}
In this paper, a model predictive controller (MPC) is proposed to generate treatment strategies for individuals with adrenal insufficiency (AI). A novel model of the HPA axis is proposed that incorporates the circadian rhythm. The proposed model is validated against patient data. An open-loop cortisol replacement strategy was used to simulate both primary and secondary AI.

An MPC-based virtual physician assistant is designed. A benefit of using MPC is the ability to impose constraints on the prescribed input doses. In a clinical trial, a negative dose of medication cannot be prescribed to a patient. Additionally, there are upper bounds on the amount of medication a person can ingest before it causes an adverse effect on the patient’s health. The MPC generated successful treatment strategies resulting in hormone levels being driven to healthy ranges. The recommended dosages were quantized to values typically available to the pharmacist.

The next stage of this research involves designing an algorithm to set the model parameters to adapt the model to patient-specific treatment and testing the controller on actual patient data to verify the prescribed medication dosing strategy. This would further refine the model and controller to produce accurate results for a specific individual. Additionally, integrating advanced machine learning algorithms, such as those presented in~\cite{chong2023introduction}, could further enhance the predictive accuracy of individualized hormone dynamics, enabling more adaptive and personalized control strategies. Beyond the standard Kalman filter, exploring advanced nonlinear observer designs could further improve the accuracy of estimating unmeasured hormone concentrations. Further exploration could involve understanding the interconnections and analyzing the mathematical modeling of the HPA axis and the hypothalamus-pituitary-thyroid (HPT) axis. Both axes interact and have some influence on each other as stated in~\cite{HPA-HPT,HPA-HPT1,HPA-HPT2}. This can lead to a more accurate modeling of stress-related disorders.
\bibliography{sn-bibliography}
\end{document}